\documentclass[aapm,mph,amsmath,amssymb,reprint]{revtex4-2}

\usepackage{graphicx}
\usepackage{dcolumn}
\usepackage{bm}
\usepackage{xcolor}

\begin{document}

\preprint{AAPM/123-QED}

\title{Quantum-enhanced cluster detection in physical images}

\author{Jason L. Pereira} \email{jason.pereira@fi.infn.it}
\affiliation{Department of Computer Science, University of York, York YO10 5GH, UK}
\affiliation{Department of Physics and Astronomy, University of Florence,
via G. Sansone 1, I-50019 Sesto Fiorentino (FI), Italy}

\author{Leonardo Banchi}
\affiliation{Department of Physics and Astronomy, University of Florence,
via G. Sansone 1, I-50019 Sesto Fiorentino (FI), Italy}
\affiliation{ INFN Sezione di Firenze, via G. Sansone 1, I-50019, Sesto Fiorentino (FI), Italy }

\author{Stefano Pirandola}
\affiliation{Department of Computer Science, University of York, York YO10 5GH, UK}

\date{\today}

\begin{abstract}
	Identifying clusters in data is an important task in many fields. In this paper, we consider situations in which data live in a physical world, so we have to first collect the images using sensors before clustering them. Using sensors enhanced by quantum entanglement, we can image surfaces more accurately than using purely classical strategies. However, it is not immediately obvious if the advantage we gain is robust enough to survive data processing steps such as clustering. It has previously been found that using quantum-enhanced sensors for imaging and pattern recognition can give an advantage for supervised learning tasks, and here we demonstrate that this advantage also holds for an unsupervised learning task, namely clustering.
\end{abstract}

\maketitle

\section{Introduction}\label{section: intro}

Pattern recognition is an important task in the field of data processing\cite{bishop2006pattern}. The goal is to recognize the presence of a pattern - or to identify which of a number of possible patterns are present - in data. Automated pattern recognition can be accomplished via machine learning. This could be either supervised or unsupervised learning. In supervised machine learning, we have a set of data points that fit the patterns that we are trying to recognize (labeled by which of the patterns they fit), and the algorithm must learn from this set how to classify previously unseen data points (i.e. how to find the pattern most closely matched by the unseen data). In unsupervised learning, the goal is to identify patterns in data without the use of sample data.

Clustering is one example of an unsupervised learning problem. The goal of clustering is to identify clusters of similar data points in a data set. Part of the problem is defining what constitutes a cluster in the first place\cite{estivill2002so}. Intuitively, a cluster should be an region of the parameter space in which there is a high density of data points with similar values, and any two clusters should be distinct from each other. Often, there will need to be a trade off between these two intuitive ideas. How this intuition is formalized depends on the type of algorithm used, and so different types of algorithm can find different clusterings.

In some settings, the data will first need to be collected before processing it\cite{genovese2016real}. For instance, suppose we have a surface that we wish to image and identify clusters on. In a microscopy setting, we may wish to find clusters of similar particles on a surface\cite{abobeih2019atomic}. Kerr microscopy is often used to identify domains in thin magnetic films\cite{soldatov2017advances}, which can be regarded as a clustering problem. In a biological setting, we may want to identify structures in cells\cite{ounkomol_label-free_2018,kandel_phase_2020}. In quantum reading\cite{pirandola_quantum_2011}, we have a data-set encoded in set of pixels with different values, and we may want to identify clustering or structure in the underlying data.

The improved accuracy offered by using quantum measurements has previously been shown to provide an advantage for supervised learning tasks\cite{banchi_quantum-enhanced_2020,harney2021ultimate}. We might therefore expect a similar advantage for unsupervised learning tasks, such as clustering. In this paper, we compare protocols that use classical measurements for data collection with protocols that use quantum measurements, in order to establish whether such an advantage exists.

We note that this is distinct from using quantum computers to implement classical machine learning algorithms\cite{biamonte_quantum_2017}, and the improvement that we are looking to prove here is in accuracy, as in quantum metrology\cite{giovannetti_advances_2011}, rather than a speedup in computation time\cite{lloyd2013quantum}. If we start with classical information, we cannot gain an accuracy advantage, because any operation that can be done by a quantum computer can be performed by a classical computer (albeit slower). The aim of gaining accuracy rather than speed is similar to Refs.~[\onlinecite{gambs_quantum_2008}] and [\onlinecite{sentis_unsupervised_2019}], which consider clustering of quantum states, rather than using quantum computers to speed up the clustering of classical states.

The accuracy advantage that we can gain is due to the improved imaging possible due to using quantum probes. Although one might intuitively expect that an improvement in the data used for clustering will lead to improved cluster detection, it is not obvious that the quantum advantage in imaging is significant enough to affect the results of a clustering algorithm, since it could be the case that a data processing step could reduce the advantage due to the better data to the point of any advantage becoming negligible. In this paper, we show that a quantum advantage in imaging is robust enough to survive a clustering algorithm.

In Section~\ref{section: task overview}, we outline the imaging and clustering task that we are considering. In Section~\ref{section: classical clustering}, we give a brief overview of classical clustering algorithms and how clusterings can be validated and compared. In Section~\ref{section: comparison}, we carry out a numerical comparison of quantum and classical protocols for specific clustering tasks by simulating them using MATLAB. Section~\ref{section: conclusion} is for conclusions.

Some details of the numerical comparison in Section~\ref{section: comparison} are found in the appendices. In Appendix~\ref{appendix: ROC}, we calculate the receiver operating characteristic used in our numerical studies for both the classical and the quantum cases. In Appendix~\ref{appendix: MI}, we explain how we calculate the mutual information values in the numerical studies. The code used to carry out the calculations and generate the plots is available as Supplemental Material\cite{supp}.

\section{Overview of the task}\label{section: task overview}

\begin{figure}[tb]
\vspace{+0.1cm}
\centering
\includegraphics[width=1\linewidth]{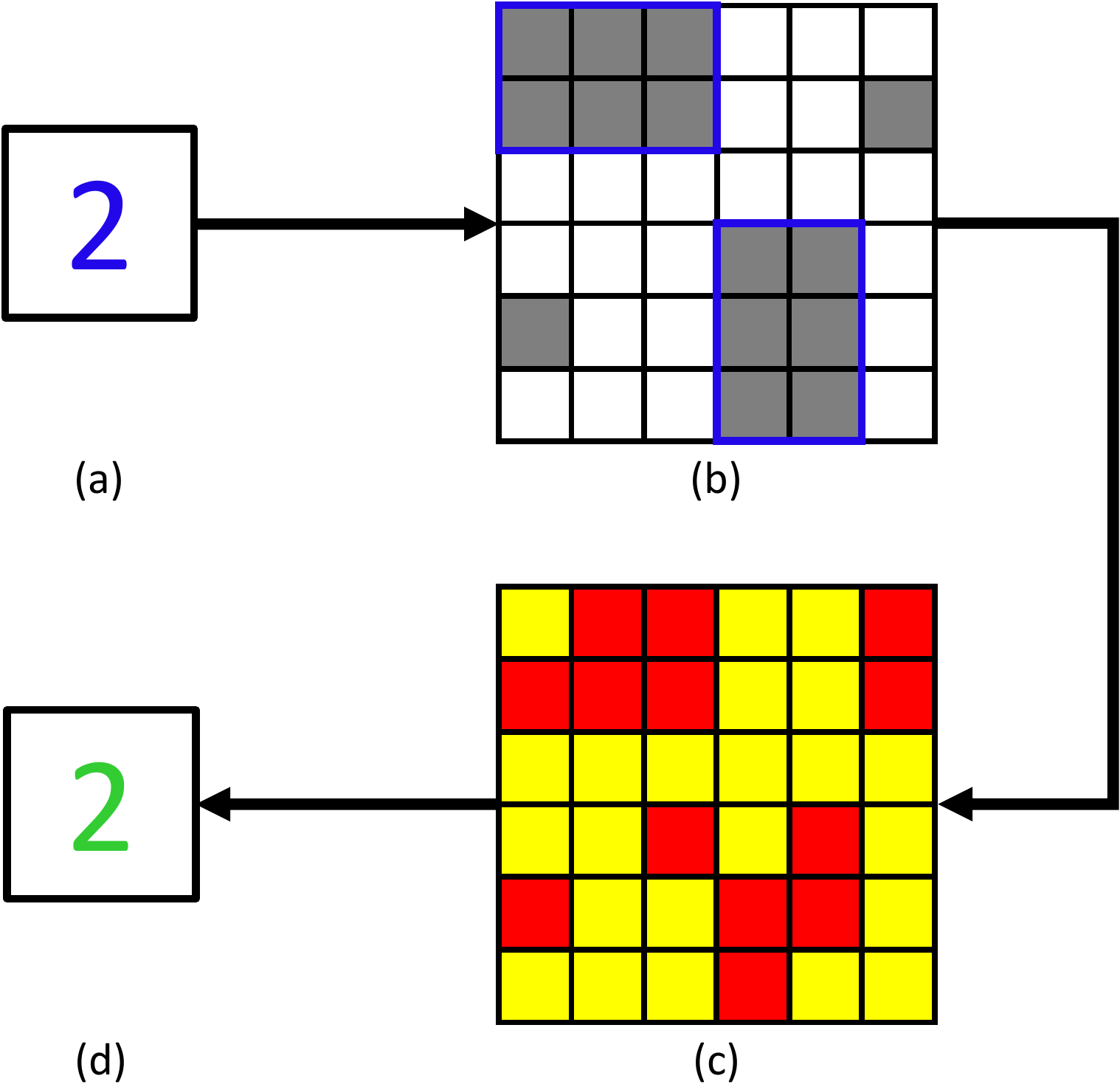}\caption{(a) The ground truth, namely the number of clusters (two, in the shown example), (b) the true channel pattern, (c) the measurement result, and (d) the result of the classical clustering (two). In the shown example, a cluster is a block of $2$ by $3$ dark pixels (which could represent collections of particles, structures, etc). The  number of clusters (a) is the only piece of information that we are interested in, rather than the position of the clusters on the surface (if we were interested in the positions, that information would also be part of the ground truth). (b) is how this ground truth corresponds to the physical multi-channel that we probe. Here, each pixel is one of two types of channel: gray, representing the presence of a particle, or white, representing the absence of particles. The two clusters are outlined in blue, but note that there are two gray pixels that are not part of either of the clusters. These could represent, for instance, particles that do not form part of one of the structures that we are looking for. If we had perfect knowledge of the state of (b), we would then carry out classical clustering algorithms in order to find the two outlined clusters and recover the value of (a). Instead, we carry out measurements on the multi-channel in order to estimate (b), with the result being (c), where red squares represent pixels we believe to be gray in (b) and yellow squares represent pixels we believe to be white. Our measurement process cannot perfectly reproduce (b), and so some squares that are white in (b) are red in (c) and some squares that are gray in (b) are yellow in (c). These are the two types of misdetection that can occur in this scenario. Finally, we carry out some classical clustering algorithm on (c) in order to estimate the number of clusters, and hence (a). This estimate is (d). Note that in this figure, the algorithm has only found two clusters, and hence has estimated (a) correctly, despite the extra red pixels. A different, potentially less appropriate, clustering algorithm might mistakenly decide that the two red pixels in the top-right corner of (c) are part of a cluster, and so overestimate (a).}
\label{fig: cluster to measurement}
\end{figure}

Let us formalize our setting by considering a surface that has been divided into a grid of pixels. We believe that the surface contains some form of underlying structure (i.e. a global property of the surface), and we are interested in identifying it. In Fig.~\ref{fig: cluster to measurement}, this ground truth is represented by (a). The ground truth could be the positions, numbers, or shapes of clusters of particles (in the setting of microscopy) or could be how the surface is divided into domains (in the setting of microscopic films), etc. It consists of everything (and only the things) that we want to find out.
In the example shown in Fig.~\ref{fig: cluster to measurement}, this is the number of $2\times3$ blocks of dark pixels. 

We mathematically represent the surface as a multi-channel, with each pixel being represented by a quantum channel, $\mathcal{C}_i$, where the label denotes the position of the pixel ((b) in Fig.~\ref{fig: cluster to measurement}). These channels represent the interactions of the pixels with any probes sent at them. Each channel is drawn from a set of possible channels (such as a set of lossy channels with different reflectivities), and which channel we have is determined by some underlying classical parameter, $\phi_i$. The identity of the channel enacted by a given pixel is determined by $\phi_i$, so that identifying the channel $\mathcal{C}_i$ is equivalent to measuring $\phi_i$. The data set $\{\phi_i\}$ is what we call the channel pattern and is what we are trying to find out by imaging the surface. For instance, the presence or absence of a particle could each correspond to one of two possible quantum channels. In this binary case of each pixel being one of only two possible channels, each $\phi_i$ will be drawn from the set $\{0,1\}$. If we have more possible channels, these could each correspond to different types of particles being present (and we would have more possible values of $\phi_i$). Alternatively, the channel could be continuously parameterized, corresponding to continuous data (for instance the magnetization vector of each pixel in a thin magnetic film).

This data set, $\{\phi_i\}$, will stochastically depend on the ground truth, but may not be entirely determined by it. There could, for instance, be unclustered particles randomly distributed over the surface. In Fig.~\ref{fig: cluster to measurement}, (a) gives the number of clusters, but not their positions or orientations. In other words, the mapping from (a) to (b) may not be deterministic. Any information present in (b) that is not present in (a) is information that a classical clustering algorithm would aim to discard.

Suppose our goal is to carry out pattern recognition on the data set, $\{\phi_i\}$, which corresponds to the channel pattern. A protocol to do so would consist of two stages: carrying out measurements to estimate the parameters $\phi_i$ and then clustering the resulting data set.

We can image the surface by sending probes to interact with each of the pixels. These probes can be regarded as quantum states, and could consist of photons or of some other type of particles, such as electrons (e.g. in electron microscopy). Some measurement is then carried out on the return states, in order to determine the quantum channels corresponding to each pixel. In other words, we carry out quantum channel tomography (or metrology if the parameterization of the possible channels is continuous) in order to reconstruct the true channel pattern ((b) in Fig.~\ref{fig: cluster to measurement}) from the measurement result ((c) in Fig.~\ref{fig: cluster to measurement}).
Note that if we are able to use unlimited energy to probe the pixels, we will be able to perfectly reconstruct the channel pattern, so there will be no difference between (b) and (c). However, if we limit the energy of the probes sent through each pixel, we will (in general) have an imperfect reconstruction of (b). This constraint could be due to our sample being sensitive, so that an energetic measurement would be destructive.
We can then carry out classical clustering algorithms on the measurement result to get (d), an estimate of the ground truth (a).

We can categorize possible protocols based on the type of probe used: specifically whether or not the probes have a positive P-representation. We call states with a positive P-representation (such as coherent states) classical, and all other states, such as two-mode squeezed vacuums (TMSVs) and number states, quantum.
Other quantum states, with multipartite entanglement, were 
considered for imaging purposes in Ref.~[\onlinecite{pereira_idler-free_2020}].

We will therefore call protocols that probe the pixels with classical states and then carry out clustering on the measurement results ``classical protocols" and will call protocols that probe the pixels with quantum states and then carry out clustering on the results ``quantum-classical protocols". This latter choice of name is to allow for a potential third type of fully quantum protocol (not considered in this paper) that probes the surface with some collective quantum state - potentially with entanglement between pixels - and then carries out a collective measurement on the return state in order to extract a global property (the ground truth), rather than probing the surface pixel-wise and carrying out classical clustering on the result. However, the quantum-classical class of protocols is more relevant from a near-term perspective, since TMSVs, which could be used as signal-idler pairs to individually probe the pixels, can be generated using current technology. This may not be the case for more general states used in fully quantum protocols.

In this paper, we will consider classical and quantum-classical protocols that send a single probe through each pixel, with the same energy constraint for each pixel, and then carry out a classical clustering algorithm on the results. Although the optimal quantum-classical protocol cannot be worse than the optimal classical protocol at reconstructing the channel pattern (recovering (b) from (c)), due to the fact that classical protocols are a special case of quantum-classical protocols, there is no guarantee that any advantage gained at this stage will be retained through the data processing. It is therefore worth investigating whether the improved imaging capabilities afforded by quantum states can lead to a non-negligible advantage in clustering accuracy.

\section{Classical clustering} \label{section: classical clustering}
\subsection{Types of clustering algorithms}

Two of the main, basic types of classical clustering algorithms are centroid-based algorithms (such as k-means\cite{lloyd_least_1982}) and density-based algorithms (such as DBSCAN\cite{ester_density-based_1996}). With centroid-based algorithms, the goal is to minimize the sum of the distances (via some metric) of the points to the nearest centroid, for a fixed number of centroids. For density-based algorithms, we grow clusters as areas with a high density of points (and some density cut-off). Other possible types of algorithms include distribution-based methods, in which we fit our points to some distribution (e.g. a sum of Gaussians) where parameters of the distributions are unknown. Each method may correspond to a different scenario we may be interested in and the ``best" type depends on the specific setting.

In k-means, we must choose the number of clusters in advance (as $k$) and then we find the $k$ centroids (points) such that the total distance from each point to the nearest centroid is minimized. Distance may be Euclidean, squared-Euclidean, Bures, etc. By design, the clusters will always be roughly spherical (in terms of the distance metric). k-means assigns every point to a cluster. One common variant of k-means is k-medoids\cite{kaufman_partitioning_1990}, which is similar to k-means, but limits the possible centroids to the data points themselves (rather than allowing them to be located anywhere in the coordinate space).

In DBSCAN, we calculate the density of points in a region around each point (the size of the region, $\epsilon$, is a parameter for the algorithm). If this is above a cut-off, the point is a core point of a cluster. If not, it is either a non-core point or not in a cluster. Such algorithms find all clusters rather than specifying the number in advance. DBSCAN may decide that some points are not part of a cluster at all.

Whilst both k-means and DBSCAN assign each data point to at most a single cluster, variants exist that carry out ``fuzzy" clustering, in which data points can be assigned to multiple clusters, with different degrees of affiliation.

\begin{figure}[tb]
\vspace{+0.1cm}
\centering
\includegraphics[width=1\linewidth]{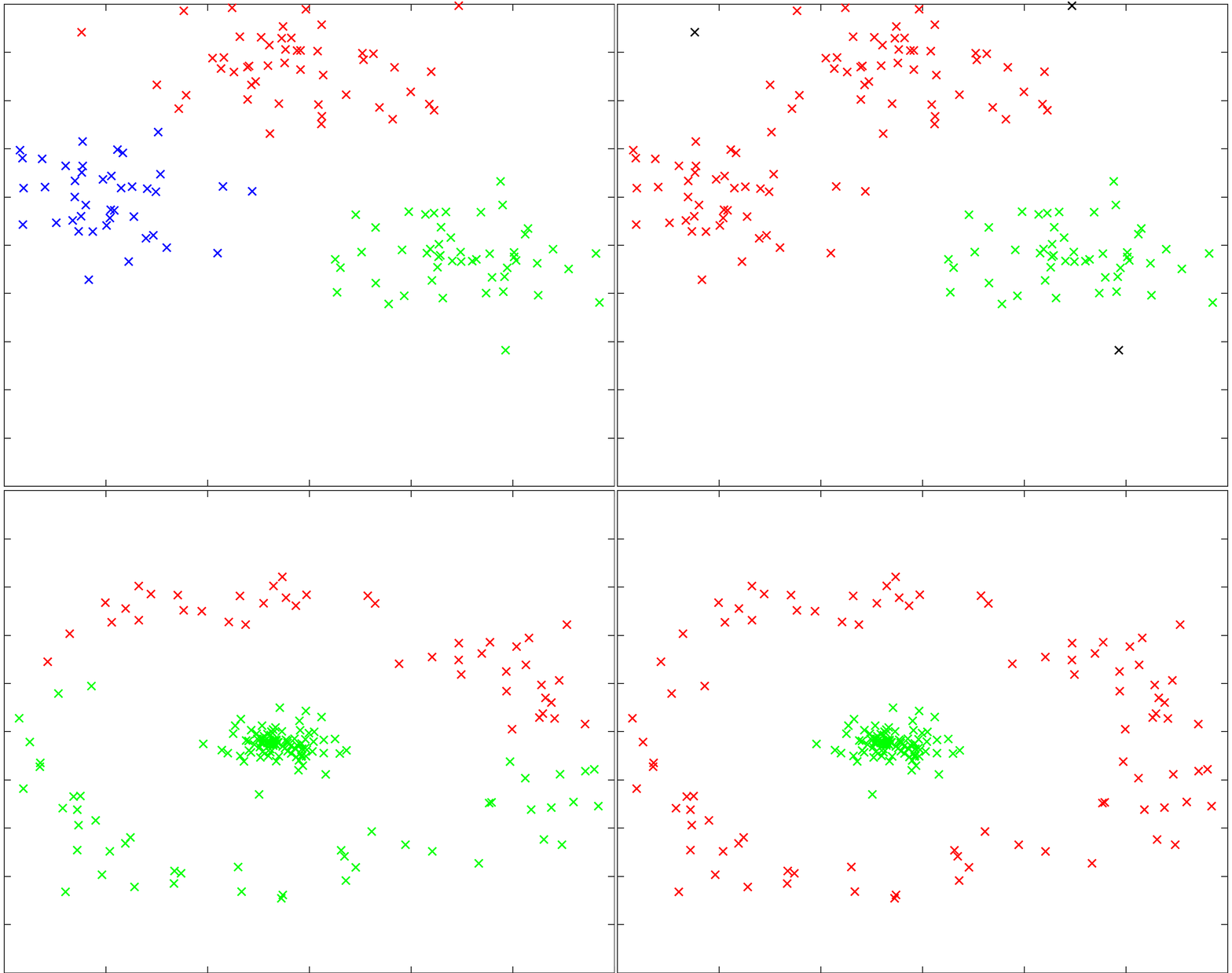}\caption{k-means clustering (left) and DBSCAN (right) on two different data sets. The top data set was generated as the sum of three normal distributions with different mean values, whilst the bottom data set was generated as a normal distribution with a surrounding ring. k-means clustering (with three clusters) is better than DBSCAN at identifying the three (roughly circular) clusters corresponding to each of the normal distributions in the top image, whilst DBSCAN is better (than k-means clustering with two clusters) at identifying the structure of a ring with a circle inside in the bottom image. Note that only the DBSCAN images have unclustered points (in black).}
\label{fig: clusterings}
\end{figure}

Which type of algorithm is appropriate depends on our setting and what we want to discover. This relates to the previously mentioned question of what constitutes a cluster. For supervised learning, we have a correct answer and can calculate the error. For unsupervised learning, we only have intuitive ideas of what a good clustering ``should" look like.

Consider Fig.~\ref{fig: clusterings}, which compares k-means clustering and DBSCAN for two different data sets. The top data set was generated by picking three pairs of coordinates and then choosing coordinates randomly from normal distributions centered on those three points, whilst the bottom data set was generated by picking a pair of coordinates and picking points from both a normal distribution centered on it and from a ring around it. k-means clustering with three clusters is able to identify three distinct, roughly circular sets of points in the top data set, whilst DBSCAN groups two of these sets together, due to them having some overlap. In the bottom data set, DBSCAN is able to separate out the ring and the circle in the middle and identify both as being clusters whilst k-means clustering is inherently unable to identify one cluster inside another, since both clusters have the same center, and so simply splits the ring in two. In each case, one of the clustering techniques correctly identifies the ground truth - the underlying distribution used to generate the data set - whilst the other does not. The appropriate clustering algorithm depends on the data itself and what information we want to extract from it.

\subsection{Cluster validation}

How can we assess how well clustering has been done? There are several metrics that we can use, each of which weights the properties that we want for our clusters differently. Methods for assessing a clustering can be broadly divided into two categories: internal and external.

An internal metric compares a clustering to our intuitive idea of what a cluster should be. Clusters should be compact (small intra cluster distances) and well separated (large inter cluster distances). One possible metric is the sum of the distances between points and the center of their cluster. By definition, this is minimized by k-means (so long as it finds a global, rather than a local, minimum), but there is no guarantee that clusters will be well separated. It also performs poorly (compared to our intuitive idea of what clusters look like) when the clusters in the data are non-spherical. Other options include the Dunn index, which takes into account both the intra and the inter cluster distances, and the silhouette coefficient.

An external assessment assumes the existence of a ground truth. In other words, there exists some objectively correct clustering that we need to find. In a physical imaging context, this will often be the case: we carry out imaging in order to determine some real property of interest about the surface we are imaging. To assess our closeness to the true clustering, however, we would need to know the ground truth beforehand. If our goal is to find the ground truth, this can present a problem, since we would not be able to use it to assess the validity of our clustering. However, external validation can be of use in comparing different imaging and clustering protocols, to see which is better at extracting the information we want about the underlying structure.

\subsection{Comparing clusterings}

If we want to assess how different measurement techniques affect the clustering found, how can we go about doing this? If we were to use an internal validation method, we might find that a less good measurement method (one that less faithfully reconstructs $\{\phi_i\}$) results in better clustering performance than a better measurement method. However, this would simply be telling us that the measurement data produced by the worse measurement is more clustered than the data produced by the better measurement, and would not give any information about which method is closer to the ``true" clustering.

Instead, we can use an external validation method. Specifically, we can compare the clustering resulting from a given protocol to the ground truth clustering. We can do this in terms of the mutual information between the ground truth and the estimation of the ground truth that we get from our measurement result. The mutual information is a classical quantity that tells us how much information the measurement result ((c) in Fig.~\ref{fig: cluster to measurement}) holds about the ground truth that we are trying to discover ((a) in Fig.~\ref{fig: cluster to measurement}). The mutual information is routinely used for machine learning purposes, forms the basis of the information bottleneck method\cite{tishby1999information}, and was employed to study the generalization and classification errors in supervised quantum classifiers\cite{banchi2021generalization}. For a given protocol, we can calculate the reduction in uncertainty about the position of the clusters due to the use of the protocol - quantified by the mutual information - and compare this to the reduction in uncertainty achieved by a different protocol.

Note that the best possible measurement will not necessarily be one that maximizes the mutual information between the measurement result and the channel pattern ((b) in Fig.~\ref{fig: cluster to measurement}), since it could be the case that a protocol that discards some of the information that is not relevant to the ground truth (e.g. has a low probability of detecting unclustered points or outliers) is better at identifying the ground truth than one that more faithfully reproduces the channel pattern.

\section{Classical versus quantum-classical pattern detection}\label{section: comparison}

In both discrete variable and continuous variable settings, we can gain a quantum advantage for channel discrimination and metrology for a large variety of sets of possible channels. Since quantum probes can give a quantum advantage in discriminating between possible channels, we expect them to be able to better determine the ground truth.

The extent of the quantum advantage depends on the set of possible channels that we are considering. For instance, if we are discriminating between a Pauli-Z channel and the identity channel, an entangled state can perfectly discriminate between the two options, whereas a classical state cannot distinguish between them at all. This means that it is not possible to make general statements about the advantage offered by using quantum states, beyond the fact that they offer an improvement.

Instead, we will consider two specific examples of problems: one involving centroid-based clustering and the other involving density-based clustering. We will compare classical protocols and quantum-classical protocols, restricting ourselves to one-shot, non-adaptive protocols.

The types of problem we will consider will involve imaging surfaces divided into grids of pixels and clustering the measurement results. In the scenarios we will consider, imaging the pixels is a quantum reading type task - each of the pixels has a binary value ($0$ or $1$) and each possible value corresponds to a different lossy channel ($\mathcal{C}_0$ and $\mathcal{C}_1$)\cite{pirandola_quantum_2011}. In each case, we will consider the same pair of possible channels, but with different types of underlying channel patterns and consequently different clustering strategies after carrying out the measurements.

Since we have only two possible channels, we have two types of errors. These are the type 1 error - the probability of detecting a particle when no particle is present - and the type 2 error - the probability of not detecting a particle that is present. The plot of the achievable type 1 error for a given type 2 error, for a given detector, is called the receiver operating characteristic (ROC).

For certain pairs of channels and energy constraints, we can prove a quantum advantage in discriminating between them, meaning that we can achieve a lower type 2 error for a fixed type 1 error. Our aim is to show, via numerical simulations (in MATLAB), that the lower type 2 errors obtained by using quantum imaging protocols can lead to an increase in clustering accuracy, and hence that using quantum, rather than classical, imaging protocols can allow us to gain more information about the ground truth.

\begin{figure}[tb]
\vspace{+0.1cm}
\centering
\includegraphics[width=1\linewidth]{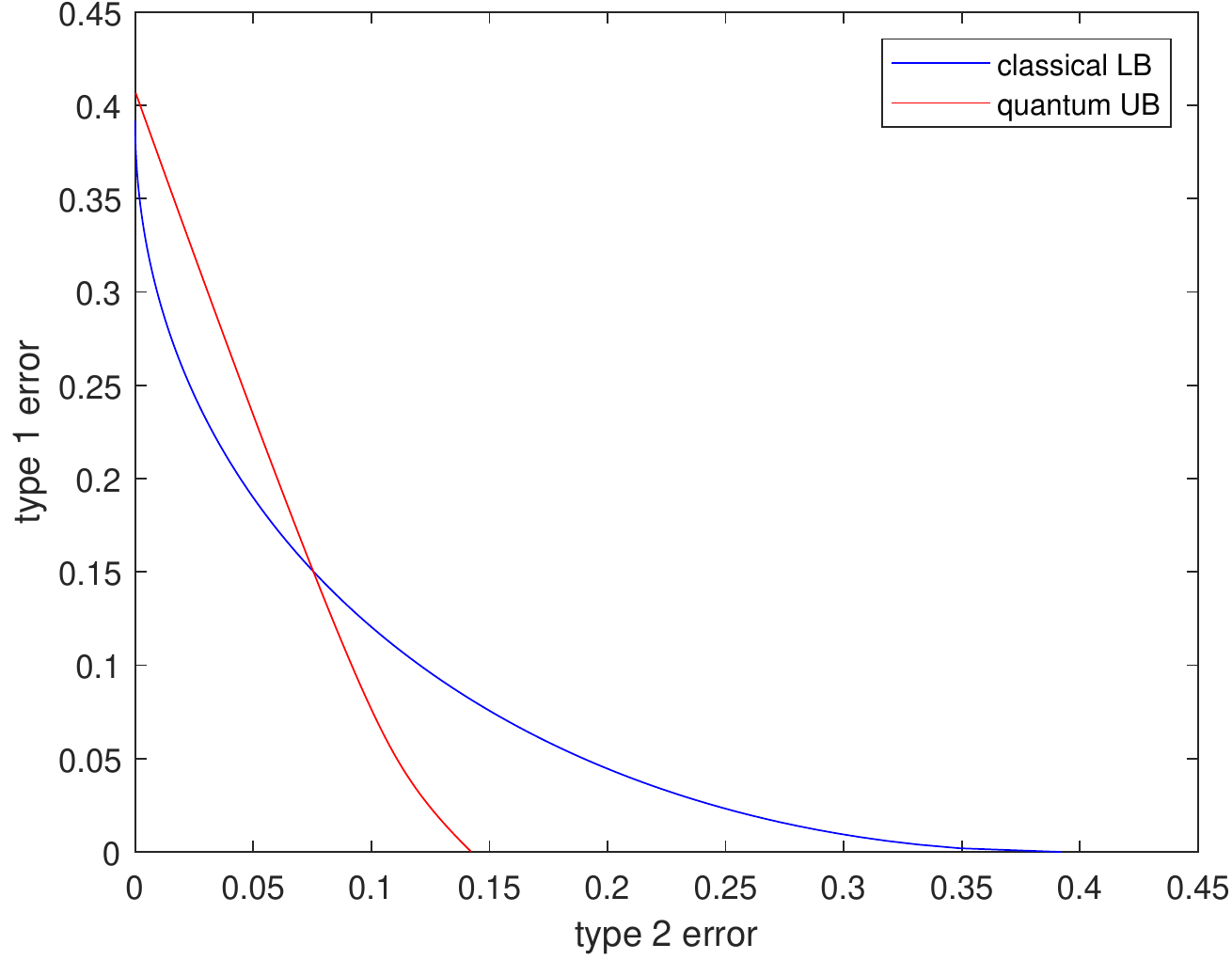}\caption{Receiver operating characteristic for classical and quantum protocols, each with an average photon number of 8, discriminating between two pure loss channels, one with a transmissivity of 0.95 and one with a transmissivity of 0.4. The blue line is the lower bound for classical protocols (protocols using probes with positive P-representations), whilst the red line is an achievable curve for quantum protocols (and therefore constitutes an upper bound on the optimal quantum protocol). For low type 1 errors, we can prove a quantum advantage in terms of measurement errors.}
\label{fig: ROC}
\end{figure}

In order to numerically compare quantum and classical protocols, we must choose some concrete parameter values. We let both channels be pure loss channels, setting the transmissivity of $\mathcal{C}_0$ to 0.95 and the transmissivity of $\mathcal{C}_1$ to 0.4. We constrain the per-pixel probe energy to a mean photon number of 8. This gives us the ROC in Fig.~\ref{fig: ROC}. See Appendix~\ref{appendix: ROC} for details of how the ROC was found for each case.

For each type of protocol, we choose pairs of type 1 and 2 errors from the ROC curve in Fig.~\ref{fig: ROC}, for type 1 error values between 0 and 0.05, and carry out simulations using these pairs of errors. The reason for using this range is that for large values of the type 1 error, the results become almost entirely random, since the number of false positives becomes comparable to the number of true positives. This means that, for a classical protocol, the type 2 error varies between approximately $0.1899$ (for a type 1 error of $0.05$) and $0.3918$ (for a type 1 error of $0$), whilst for a a quantum protocol, the type 2 error varies between approximately $0.1107$ (for a type 1 error of $0.05$) and $0.1424$ (for a type 1 error of $0$).

Note that the classical type 2 errors are lower bounds (the best possible type 2 errors for fixed type 1 errors), whilst the quantum type 2 errors are upper bounds on the best possible type 2 errors for fixed type 1 errors. Consequently, any calculated mutual information values (between the ground truth and our estimate of it) that use the classical error values will be upper bounds for classical protocols, whilst mutual information values that use the quantum error values will be lower bounds on the achievable values for optimal quantum protocols.

We will refer to the variable encoding our ground truth as $A$, the variable encoding the channel pattern as $B$, the measurement result as $C$, and the resulting estimate of the ground truth as $D$ (see also Fig.~\ref{fig: cluster to measurement}). $A$ is some global property of the surface (i.e. not a property of any particular single pixel) that we want to find out. $B$ is the actual pattern of pixels on the surface (a matrix of 0s and 1s), which is correlated to $A$. Each protocol images each pixel, using either classical or quantum states, to get a different matrix of 0s and 1s, and then carries out clustering on the result. The matrix of 0s and 1s constitutes the variable $C$ (our estimation of $B$), whilst the result of the clustering algorithm gives us the variable $D$ (our estimation of $A$). We can quantify how good a protocol is at obtaining the ground truth by finding the mutual information between variables $A$ and $D$. This can be expressed as
\begin{equation}
    I(A:D) = H(D) - H(A|D).
\end{equation}

\begin{figure}[tb]
\vspace{+0.1cm}
\centering
\includegraphics[width=1\linewidth]{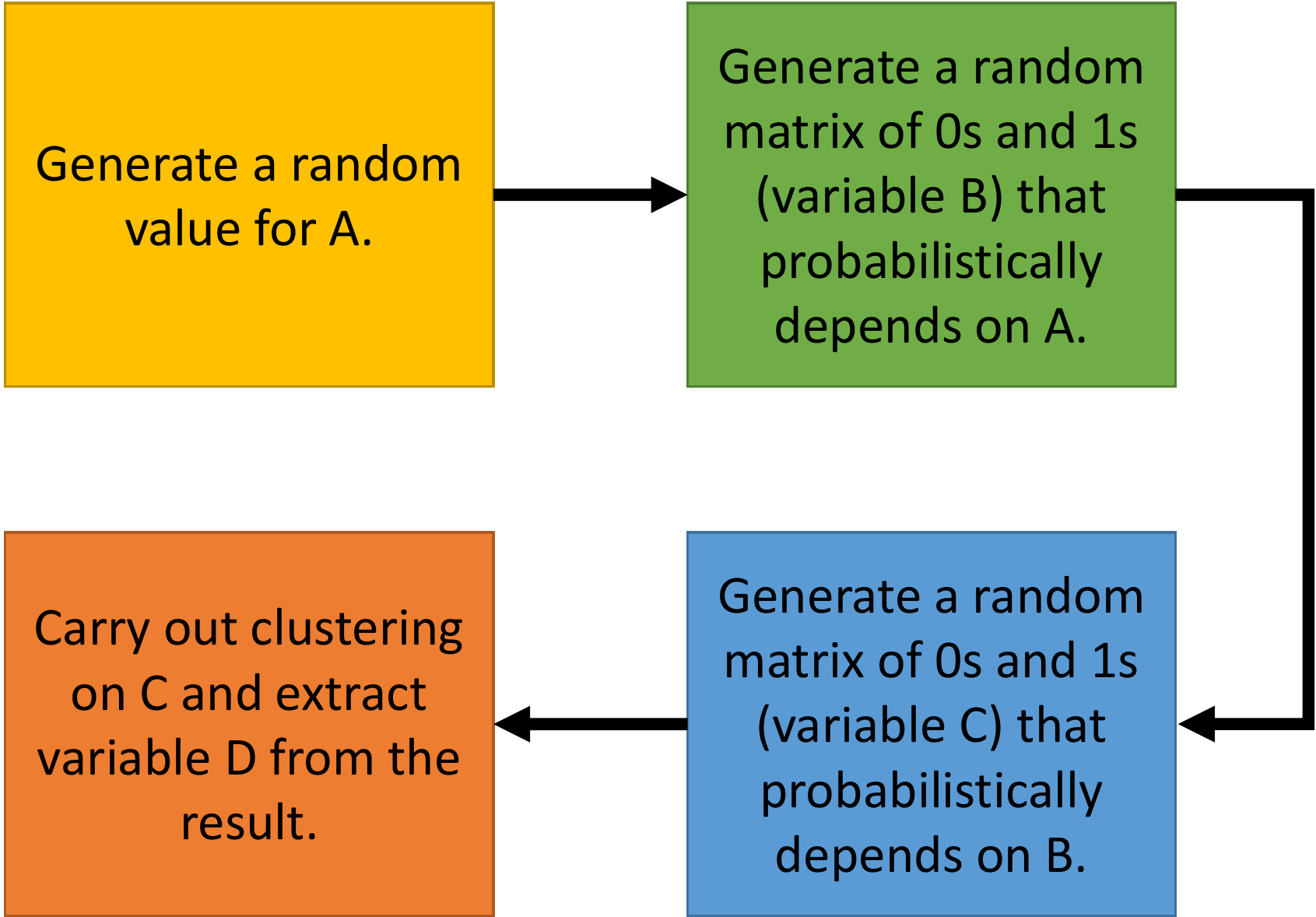}\caption{Flowchart outlining the numerical simulation process used to compare classical and quantum-classical protocols.}
\label{fig: flowchart}
\end{figure}

The full simulation process for a given pair of type 1 and 2 errors is as follows. First, randomly choose a value of the ground truth variable, $A$. Next, generate a true channel pattern (matrix of $0$s and $1$s) based on the value of $A$. This channel pattern is the variable $B$, and the probabilistic mapping between $A$ and $B$ depends on the specific scenario. Then, probabilistically generate a measurement result, $C$, based on $B$. This is again a matrix of $0$s and $1$s. If an entry in matrix $B$ is $0$, the corresponding entry in $C$ is $0$ with probability $1-\xi_1$ and is $1$ with probability $\xi_1$, where $\xi_1$ is the type 1 error. If an entry in matrix $B$ is $1$, the corresponding entry in $C$ is $0$ with probability $\xi_2$ and is $1$ with probability $1-\xi_2$, where $\xi_2$ is the type 2 error. A clustering algorithm is then carried out on variable $C$ (the choice of algorithm and the settings depend on the scenario) to generate variable $D$. Finally, variables $A$ and $D$ are recorded.

These steps are repeated a large number of times so that the mutual information between $A$ and $D$ can be estimated. The process for generating each sample is outlined in Fig.~\ref{fig: flowchart}. Note that the process is a Markov chain: the estimation of the ground truth, $D$, depends only on the measurement result, $C$, which depends only on the channel pattern, $B$, which depends only on the ground truth, $A$.

\subsection{Classical versus quantum-classical k-medoids clustering}\label{subsection: k-medoids}

In our first scenario, we have a surface on which there are a number of non-interacting particles. We divide this surface into an $d$ by $d$ grid of pixels and image each pixel to determine the presence of absence of a particle in that pixel. We assume that a pixel can be occupied by at most one particle, so that we have only two choices of possible channels, which we can label as $\mathcal{C}_0$ (no particle) and $\mathcal{C}_1$ (particle present).

\begin{figure}[tb]
\vspace{+0.1cm}
\centering
\includegraphics[width=1\linewidth]{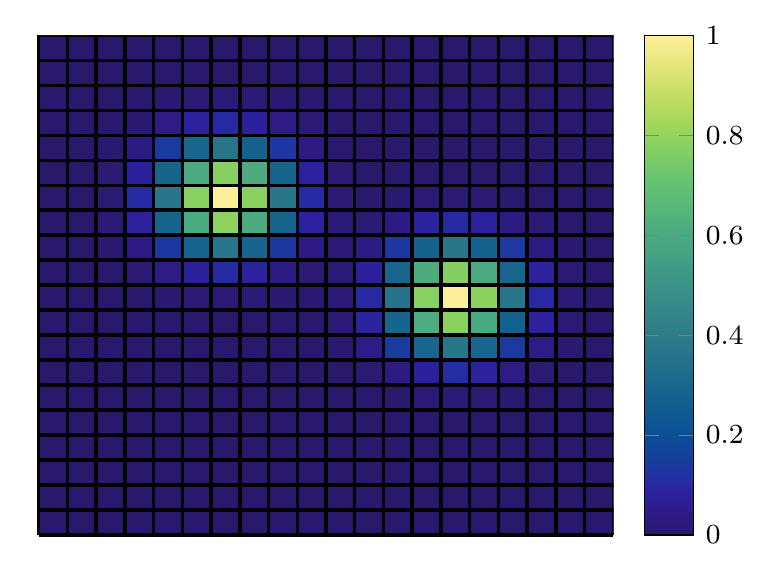}\caption{Proportion of samples in which each pixel contains a particle, for two attractors, for the scenario in Section~\ref{subsection: k-medoids}. The probability for each pixel to contain a particle for any given sample is given by Eq.~(\ref{eq: prob dist k-means}) (assessed at the center of each pixel). As can be seen, the pixels on which the attractors are centered always contain particles, whilst those further out contain particles less frequently, and those more than $4$ pixels from an attractor have a probability of containing a particle that is close to $0$. The aim of imaging the surface is to locate the two attractors.}
\label{fig: kmedoids pdist}
\end{figure}

Suppose we know that the surface contains $m$ attractors, to which the particles are attracted, so that the probability function for finding a particle near to an attractor is Gaussian. Specifically, suppose the $m$ attractors are located at coordinates $\{x_m,y_m\}$ and the probability of a pixel with its center at position $(x,y)$ containing a particle is given by
\begin{equation}
    f(x,y) = \phi \sum_{i = 1}^m e^{-\frac{(x-x_m)^2+(y-y_m)^2}{2\sigma^2}},\label{eq: prob dist k-means}
\end{equation}
where $\phi$ is some positive constant and $\sigma^2$ is some variance. For our simulations, we will set $m=2$, so that there are always exactly two attractors. The probability of finding a particle on each pixel of the surface, for two attractors and a specific choice of attractor locations, is shown in Fig.~\ref{fig: kmedoids pdist}. The probability is high close to the attractors and decays with distance from them. More than three standard deviations from either attractor, the probability of a pixel containing a particle is close to $0$. If our task is to locate the attractors, k-medoids clustering (with $m$ clusters) is a natural choice.

We emphasize that the task is to find the locations of the attractors, via their effect on the distribution of particles, rather than the locations of the particles themselves. The grid may contain any number of particles, but will always contain exactly two attractors. Furthermore, we are not able to directly detect the attractors, we can only detect the presence/absence of particles, the probability of which (for each pixel) is affected by the positions of the attractors. The purpose of this is to relate the task to a physical scenario in which clustering of the results is necessary. For instance, the attractors could be electrical charges on the surface to which the particles are attracted.

\begin{figure}[tb]
\vspace{+0.1cm}
\centering
\includegraphics[width=1\linewidth]{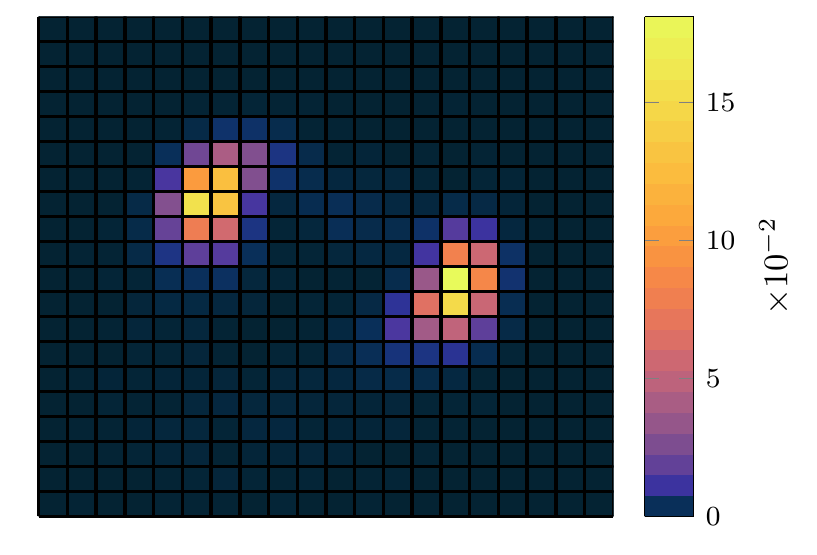}\caption{Proportion of ($10000$) samples in which each pixel is predicted to contain an attractor (i.e. to be the center of a cluster), for the scenario in Section~\ref{subsection: k-medoids}. The predicted locations follow a distribution that has a high probability close to the true locations and a low probability elsewhere. The less spread out this distribution, the less uncertainty we have about the locations of the attractors, and hence the better the imaging protocol. This measurement outcome corresponds to a type 1 error of $0.05$ and a type 2 error of $0.2$. Note that this plot does not show the same thing as Fig.~\ref{fig: kmedoids pdist}; Fig.~\ref{fig: kmedoids pdist} shows the proportion of samples in which each pixel contains a particle (and multiple pixels will contain a particle in each sample), whilst this plot shows the proportion of samples in which each pixel is found to be a cluster center (and each sample contains only, and exactly, $2$ cluster centers).}
\label{fig: kmedoids mdist}
\end{figure}

The simulation is done as per Fig.~\ref{fig: flowchart}. The coordinates of the attractors on the grid of pixels for each sample constitute the variable $A$ for that sample, and the positions of all of the particles constitute the variable $B$. The coordinates of the centroids found by the k-medoids algorithm constitute variable $D$, our estimate of $A$.

We choose random pairs of well-separated coordinates for each sample of $A$ and generate the corresponding probabilities for each pixel to have a value of $1$ in the channel pattern (variable $B$). Each value of $B$ is a matrix of $0$s and $1$s, with a probability of each that is different for every pixel and depends on $A$ (given by Eq.~(\ref{eq: prob dist k-means})). If we were to draw a large number of sample values of $B$ for a fixed value of $A$, we would get a distribution similar to  Fig.~\ref{fig: kmedoids pdist}.

As can be seen, some pixels have a value of $0$ in almost all of the samples (corresponding to the absence of a particle in that pixel), whilst some almost always have a value of $1$ (corresponding to the presence of a particle). Note that if we were to draw different values of $A$ for each sample, we would have a roughly flat distribution (although not entirely uniform, due to our constraints on the possible values of $A$, which introduce some structure).

Each value of $C$ is also a matrix of $0$s and $1$s. The probability of a pixel having each value depends on its value in the channel pattern and the two error values. Carrying out k-medoids clustering, with $2$ clusters, on $C$ results in a pair of pair of coordinates, similar to $A$.

In Fig.~\ref{fig: kmedoids mdist}, we fixed $A$ (to be the same as for Fig.~\ref{fig: kmedoids pdist}) and then randomly drew $B$ $10000$ times, then drew $C$ once for each value of $B$ (using a type 1 error of $0.05$ and a type 2 error of $0.2$), and then calculated $D$ for each value of $C$. This gave us $10000$ samples of the calculated cluster centers for a specific, fixed value of $A$. We then plotted the proportions of the samples in which each pixel contained one of the calculated cluster centers.

If the protocol were ideal, these calculated cluster centers would always be at the coordinates given by $A$, and so the positions of the cluster centers would be a perfect predictor of the positions of the attractors (the value of $A$). Fig.~\ref{fig: kmedoids mdist} shows that the actual cluster centers found are spread over a small region around the true positions of the attractors, meaning that $D$ has some entropy, even for fixed $A$, although the fact that most of the locations of the cluster centers are close to the true positions means that the protocol does gain some information about them. For further details about the calculation of the mutual information between variables $A$ and $D$, see Appendix~\ref{appendix: MI}.

To simplify our comparison of the protocols, we make the following assumptions: the attractors are well-separated from each other (i.e. the distance between each attractor and its nearest neighbor is much greater than $\sigma$), none of the attractors are close to the edge of the surface (i.e. the distance between each attractor and the nearest edge is much greater than $\sigma$), and the attractors are each located at the center of a pixel. The first two conditions simplify the probability function, whilst the third allows us to use the discrete Shannon entropy, rather than the Shannon entropy for continuous variables. The first two conditions are likely to hold if $d \gg \sigma$. We set $d=20$, $\phi=1$, and $\sigma^2=2$. We then simulate measurement results for pairs of type 1 and type 2 errors and hence calculate the mutual information for each pair of errors.

For any given type 1 error, a higher type 2 error means fewer of the particles will be detected. Conversely, a higher type 1 error means pixels will be found to contain particles when they actually do not. Both types of error are expected to reduce the accuracy with which we can estimate the cluster centers, since in one case we have less information and in the other case we have misleading information. We would therefore expect that a lower value of either error would result in Fig.~\ref{fig: kmedoids mdist} having a sharper (less spread out) distribution.

\begin{figure}[tb]
\vspace{+0.1cm}
\centering
\includegraphics[width=1\linewidth]{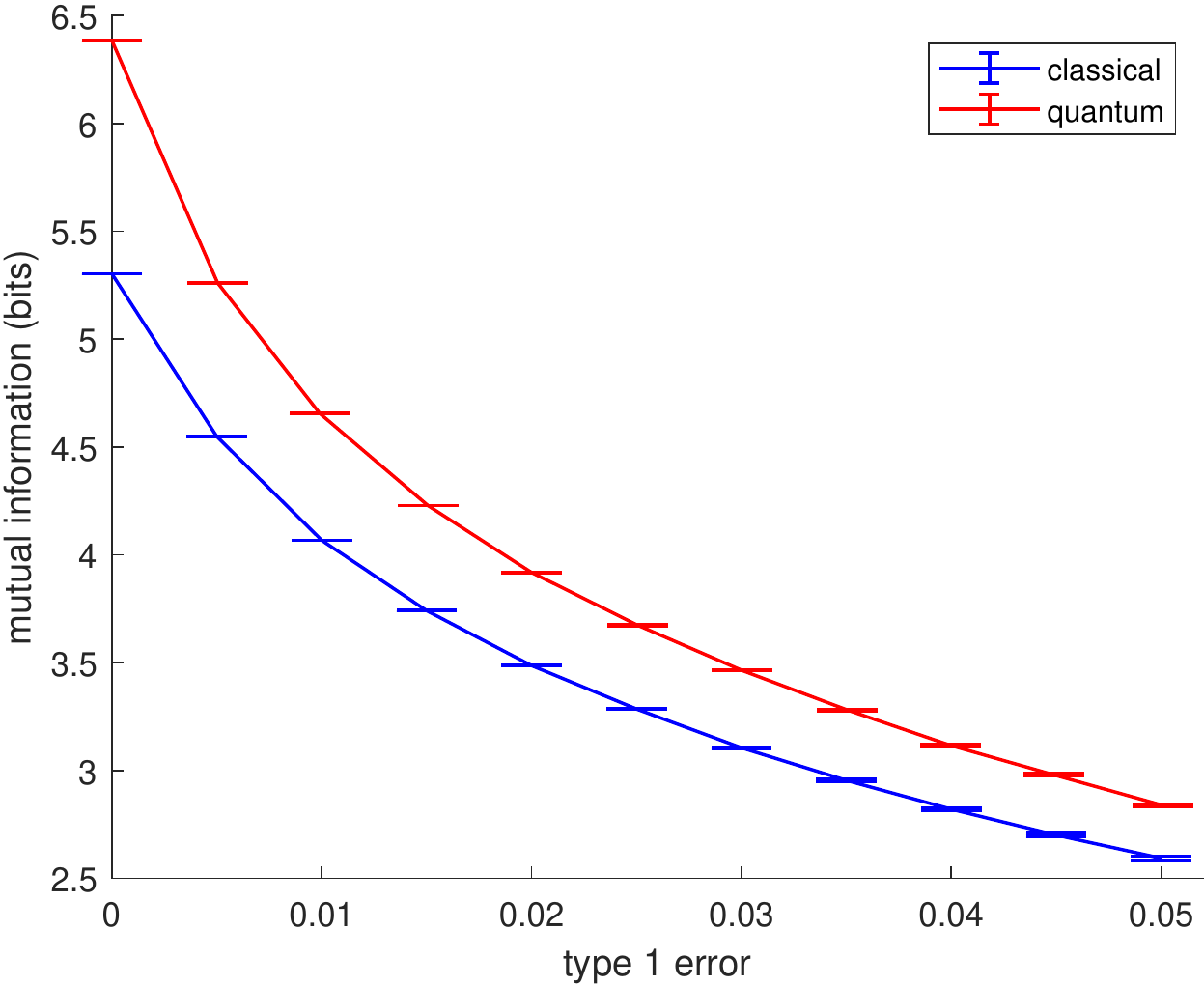}\caption{Mutual information between the ground truth and the estimated result for a pattern detection task involving k-medoids clustering, for both classical and quantum measurements of pixels. For each value of the type 1 error, measurement results are simulated for the corresponding type 2 error, for each type of measurement, and k-medoids clustering is carried out on the results. The error bars show the variance of our estimator of the mutual information.}
\label{fig: kmedoids MI}
\end{figure}

Fig.~\ref{fig: kmedoids MI} shows, as expected, that by measuring the pixels with quantum, rather than classical states, we can achieve a higher mutual information with the ground truth, and therefore can gain more information about the positions of the cluster centers. Recall that, for each type 1 error, the type 2 error for each type of protocol is given by the curves in Fig.~\ref{fig: ROC}, and that the classical type 2 error varies between approximately $0.1899$ and $0.3918$, whilst the quantum type 2 error varies between approximately $0.1107$ and $0.1424$. Note that the mutual information decreases as the type 1 error increases, for both types of protocol, despite the fact that the type 2 error decreases as the type 1 error increases.

\subsection{Classical versus quantum-classical DBSCAN clustering}\label{subsection: dbscan}

In the second scenario, we are again imaging a surface that can be represented as a grid of pixels. Suppose this surface has long, thin (non-circular) particles on it, that cover multiple pixels. For simplicity, we will assume that the particles are rectangular, with integer dimensions (in terms of pixels covered) $d_1$ and $d_2$, and that they are oriented in one of two ways (vertically or horizontally). The particles are distributed randomly over the surface, with the only constraints being that their corners lie at the corners of pixels (so that every pixel is either completely covered or not covered) and that they do not overlap.

Our task is to determine the number of such particles, which is randomly chosen from a uniform distribution between $0$ and a maximum number, $m$. DBSCAN may be more suitable than k-means clustering for identifying clusters corresponding to the particles, due to the non-circular shape of the particles and the fact that the number is not fixed or known beforehand.

\begin{figure}[tb]
\vspace{+0.1cm}
\centering
\includegraphics[width=1\linewidth]{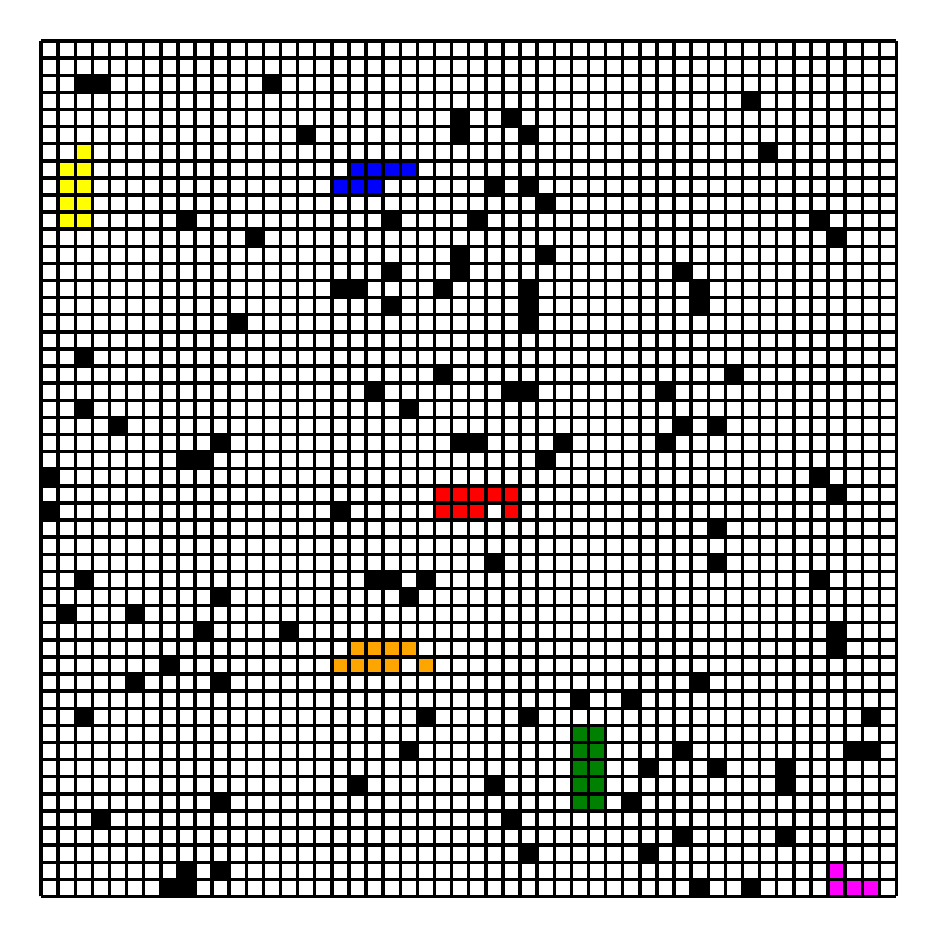}\caption{Example of a measurement result for the scenario in Section~\ref{subsection: dbscan}. Every pixel has been individually imaged and found to have a value of either $0$ or $1$ (each corresponding to a different possible channel). The particles we are looking for are $5$ pixels long and $2$ pixels wide. All of the pixels containing a particle should have the value $1$ and all other pixels should have a value of $0$. However, sometimes, we will determine a pixel to have value $0$ when its actual value is $1$ (type 2 error). Similarly, sometimes we will get a value of $1$ for a pixel that does not contain a particle (type 1 error). In order to find which pixels that we determine to have value $1$ actually contain particles and which do not, we carry out DBSCAN on the measurement results to find clusters of $1$s and only consider $1$s that are part of clusters to be real particles rather than measurement errors. All pixels found to have a value of $0$ are colored white. Pixels (found to have a  value of $1$) that are determined to belong to the same cluster are given the same (non-black) color, whilst the remaining, unclustered pixels are colored black.}
\label{fig: DBSCAN example}
\end{figure}

We again carry out the numerical simulations as per Fig.~\ref{fig: flowchart}. The number of particles present is variable $A$, the ground truth that we want to discover. For each sample, this is drawn randomly from the uniform distribution between $0$ and $m$. For a given value of $A$, we then randomly place the clusters on the grid of pixels to generate variable $B$, the channel pattern (with the constraint that no two particles overlap). After generating a measurement result (variable $C$) based on the channel pattern and the type 1 and 2 error values, we carry out DBSCAN on the measurement result to try and identify the particles. We use the number of clusters identified by DBSCAN as variable $D$, our estimate of variable $A$. For further details about the calculation of the mutual information between variables $A$ and $D$, see Appendix~\ref{appendix: MI}.

We set the number of pixels to $50$ by $50$, $m$ to $10$, and the dimensions of the particles to $2$ by $5$ pixels. For the DBSCAN algorithm, we set the minimum number of points in a region surrounding a point to identify the central point as a core point to $4$, and we set the radius of the region to $\sqrt{2}$, so that it includes all neighboring pixels (including diagonal neighbors).

Fig.~\ref{fig: DBSCAN example} shows an example measurement result for a type 1 error of $0.05$ and a type 2 error of $0.2$. The colored pixels are those identified as being part of a cluster, and hence as containing a particle, whilst the black pixels are those that have a value of $1$ in the measurement result but are not part of a cluster (and hence are assumed to be false positives).

In this example, we identify $6$ clusters, and hence this is our estimate of the number of particles. However, there are actually only $5$ particles; all of the pixels in the magenta cluster at the bottom right of the image are false positives. This shows how a high type 1 error can lead to the protocol identifying clusters that do not exist, and hence overestimating the number of particles present. Similarly, a high type 2 error could lead to a cluster not being identified at all, and hence to the protocol underestimating the number of clusters. Type 2 errors can also lead to overestimations, if the pixels falsely detected to have a value of $0$ lie in the middle of a particle. In this case, the two ends of the particle may be incorrectly found to be separate clusters. Both types of error can therefore lead to a misestimation of the number of particles, and hence will reduce the mutual information between $A$ and $D$.

\begin{figure}[tb]
\vspace{+0.1cm}
\centering
\includegraphics[width=1\linewidth]{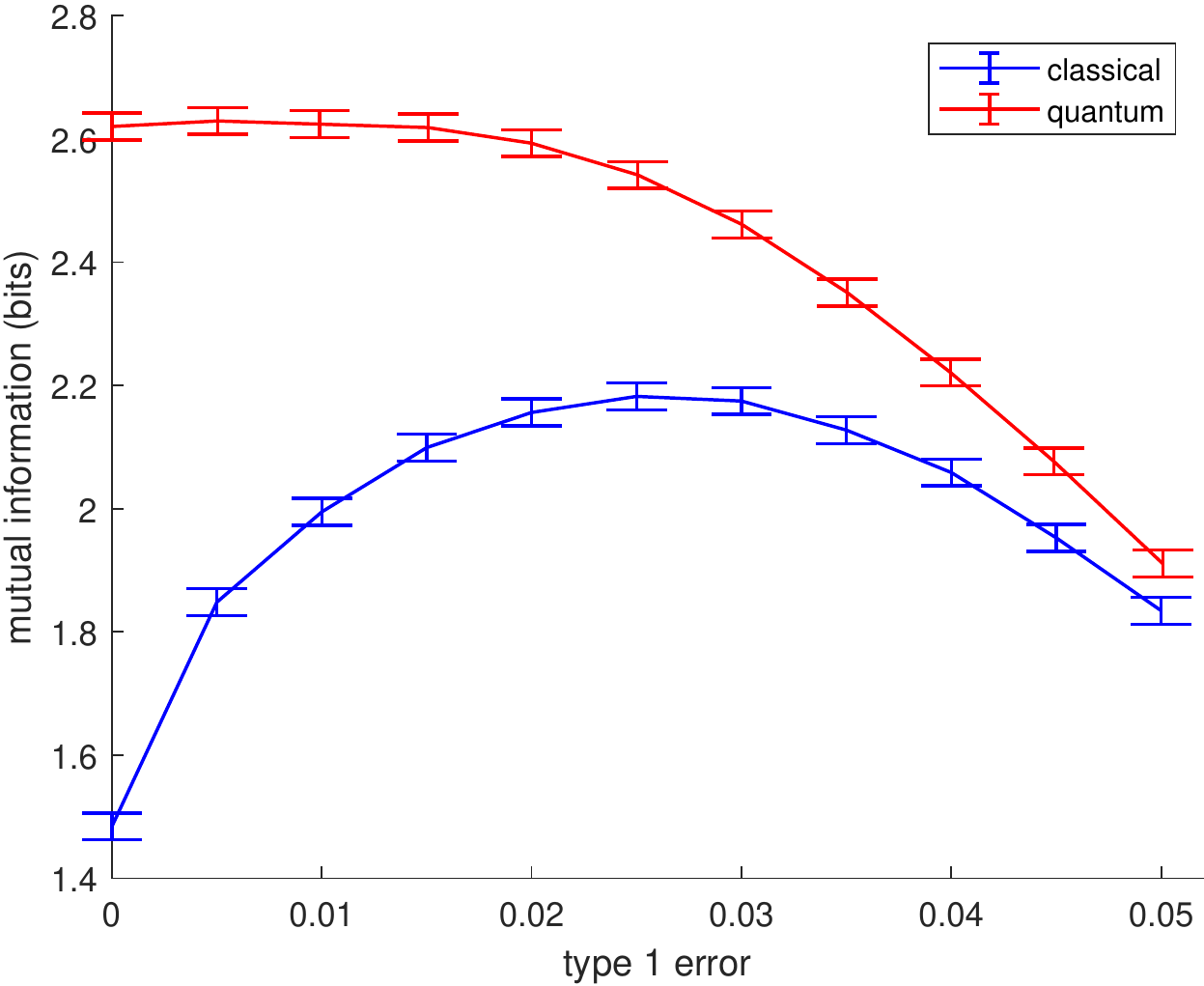}\caption{Mutual information between the ground truth and the estimated result for a pattern detection task involving DBSCAN, for both classical and quantum measurements of pixels. For each value of the type 1 error, measurement results are obtained for the corresponding type 2 error, for each type of measurement, and DBSCAN is carried out on the results. The error bars show the variance of our estimator of the mutual information.}
\label{fig: DBSCAN MI}
\end{figure}

Fig.~\ref{fig: DBSCAN MI} again shows that measurements with entangled states achieve a higher mutual information, and are therefore better at determining the number of particles, than classical measurements. For both types of protocols, the mutual information initially rises as the type 1 error increases before falling again. This is not unexpected because the type 2 error decreases as the type 1 error increases (see Fig~\ref{fig: ROC}), so the increase in mutual information is not because of the higher type 1 error but because of the lower type 2 error.

In particular, for low values of the type 1 error, we have a clear quantum advantage (for a type 1 error of $0$, the advantage is more than one bit). The large gap between the protocols, relative to their absolute values, shows that the advantage in the misdetection probability due to using quantum probes can lead to improvements in clustering accuracy for certain unsupervised learning tasks. Note that low values of the type 1 error correspond to larger differences in the type 2 error between the two types of protocols (see Fig.~\ref{fig: ROC}), although there was no guarantee that this would translate to a higher quantum advantage for lower values of the type 1 error.

\section{Conclusion}\label{section: conclusion}

Quantum states can often give an advantage over classical states when it comes to quantum imaging tasks. In this paper, we have shown that this advantage can translate to an improvement in clustering accuracy when classical clustering algorithms are performed on the results of a measurement. We demonstrated this advantage numerically for both a scenario involving the k-medoids algorithm (a centroid-based algorithm) and one involving DBSCAN (a density-based method). In both cases, using a quantum measurement, and therefore achieving a lower type 2 error for the same type 1 error, resulted in a higher mutual information between the ground truth that we wanted to know and our estimation of the ground truth based on the measurement protocol.

Whilst this result is intuitive, a small advantage in misdetection probability could easily have been lost or made negligible during the data-processing stage. It is therefore encouraging that the quantum advantage for measurements is robust enough to survive the clustering algorithms. This complements existing results showing that imaging using quantum states can give improved results for supervised learning.

Moreover, we found that different clustering algorithms and different scenarios result in different levels of quantum advantage, even for the same pairs of possible channels. Comparing Figs.~\ref{fig: kmedoids MI} and \ref{fig: DBSCAN MI} shows that the same quantum advantage in measurements can lead to different quantum advantages in clustering accuracy when applied to different problems involving different clustering techniques. Indeed, the two graphs are qualitatively different, as well as quantitatively: in Fig.~\ref{fig: kmedoids MI}, the mutual information decreases as the type 1 error increases for both the classical and the quantum-classical cases, whilst in Fig.~\ref{fig: DBSCAN MI}, both mutual informations initially peak before decreasing again. This demonstrates that the question raised in this paper is non-trivial.

Since ground truths such as the number of clusters on a surface are global properties of the entire surface, future research could consider the possibility that fully quantum protocols that collectively probe the pixel pattern as a whole, rather than individually probing each pixel, could give a further advantage over quantum-classical protocols.

Another possible consideration is protocols that involve multiple rounds of sending probes through the pixels to image them before the clustering stage. Such protocols could potentially also be adaptive, meaning that the probes used in subsequent rounds could depend on the results of measuring the return states from previous rounds. Adaptivity is known to provide an advantage for certain channel discrimination problems.

Further research could also consider situations in which there is a global constraint on the energy used to probe the entire surface, but no firm constraint on the per-pixel energy used (i.e. the sum of the energies of all of the probes is fixed, but not the distribution over the pixels).

\smallskip
\begin{acknowledgments}
J.~L.~P. and S.~P acknowledge funding from the European Union's Horizon 2020 Research and Innovation Action under grant agreement No. 862644 (FET-OPEN project: Quantum readout techniques and technologies, QUARTET). L.~B. acknowledges funding from the U.S. Department of Energy, Office of Science, National
Quantum Information Science Research Centers, Superconducting Quantum Materials and Systems Center
(SQMS) under the contract No. DE-AC02-07CH11359.
\end{acknowledgments}

\bigskip

\appendix

\section{Calculating the receiver operating characteristic}\label{appendix: ROC}

Suppose we want to discriminate between a pair of pure loss channels, $\mathcal{L}_1$ and $\mathcal{L}_2$, with transmissions of $\tau_1$ and $\tau_2$ respectively, by probing the target channel with either a classical probe or a quantum probe. In both cases, we require that the average photon number of the part of the state that passes through the channel is $\leq m$. For the classical case, we will optimize over all classical mixtures of coherent states, and therefore find the exact optimum ROC. For the quantum case, we will consider a two-mode squeezed vacuum (TMSV) probe, and will therefore find an achievable upper bound on the optimum.

\subsection{Classical ROC}

First, let us consider the output of a single coherent state sent through a pure loss channel with a transmission of $\tau$. Let $\left|\sqrt{n}\right>$ be a coherent state with an average photon number of $n$. To fully specify such a state, we should also give a phase, but the phase is not important for our calculations. Sending such a state through the channel results in the pure output state $\left|\sqrt{\tau n}\right>$. The fidelity between the outputs of channels $\mathcal{L}_1$ and $\mathcal{L}_2$ is\cite{weedbrook_gaussian_2012}
\begin{equation}
    F_{\mathrm{coh}} = \chi^n,~~\chi=\exp\left[-\frac{1}{2}|\sqrt{\tau_1}-\sqrt{\tau_2}|^2\right].
\end{equation}

Classical states can be expressed as classical mixtures of coherent states. From the linearity of quantum channels and the joint concavity of the fidelity, we can write
\begin{equation}
    F_{\mathrm{mix}} \geq f(p(n)) = \int_{n=0}^{\infty} p(n)^2 \chi^n dn,
\end{equation}
where $p(n)^2$ is the probability density function for input states with an average photon number of $n$ (and we have squared $p(n)$ to enforce non-negativity). We would like to minimize the functional $f(p)$ subject to
\begin{equation}
    \int_{n=0}^{\infty} p(n)^2 dn = 1,
\end{equation}
the normalization condition, and
\begin{equation}
    \int_{n=0}^{\infty} p(n)^2 n dn = m,
\end{equation}
the energy constraint.

We define the Lagrangian functional
\begin{align}
    &\mathcal{L}(p(n),\mu,\lambda) = \int_{n=0}^{\infty} L(p,\mu,\lambda)dn - (\mu + \lambda m),\\
    &L(p(n),\mu,\lambda) = p(n)^2 (\chi^n + \mu + \lambda n),
\end{align}
where $\mu$ and $\lambda$ are Lagrange multipliers. For a function to be a minimum of the constrained optimization, it must be a stationary point of the Lagrangian functional (a point is here defined as a specific choice of $p(n)$, $\mu$, and $\lambda$).

The stationary points of a functional are those functions for which the functional derivative is $0$. The functional derivative of $L$ is
\begin{equation}
    \frac{\partial L}{\partial p(n)} = 2p(n)(\chi^n + \mu + \lambda n),
\end{equation}
Since $\mu$ and $\lambda$ are numbers, rather than functions of $n$, we can only set $\chi^n + \mu + \lambda n=0$ for at most two different values of $n$ (since it is a continuous function of $n$ with at most one turning point). To set $\frac{dL}{dp(n)}=0$, we must have $p(n)=0$ for all but two (or fewer) values of $n$ (except in the trivial case of $\tau_1 = \tau_2$). Applying the continuity condition and the energy constraint, the candidate solutions take the form
\begin{equation}
    \begin{split}
        &p_{n_0,p_0}(n)^2 = p_0 \delta_{n,n_0} + (1-p_0) \delta_{n,n_0+\Delta(n_0,p_0)},\\
        &\Delta(n_0,p_0) = \frac{m-n_0}{1-p_0},~~0\leq p_0 < 1,~~0\leq n_0 \leq m
    \end{split}
\end{equation}
where $\delta$ is the Kronecker delta function.

We now minimize $f(p)$ over our candidate solutions. We calculate
\begin{equation}
    f(p_{n_0,p_0}) = \chi^{n_0}\left(p_0 + (1-p_0)\chi^{\Delta(n_0,p_0)}\right),
\end{equation}
and partially differentiating with regard to $m_0$, we get
\begin{equation}
    \frac{\partial f}{\partial n_0} = p_0 \ln[\chi] \chi^{n_0} \left(1-\chi^{\Delta(n_0,p_0)}\right).
\end{equation}
This is always negative if $p_0>0$ and $n_0<m$, so to minimize we can set $n_0=m$. The minimum fidelity can therefore be achieved by using a pure coherent state with an average photon number of $m$. The minimum fidelity between output states for a classical probe is
\begin{equation}
    F_{\mathrm{class}} = \chi^{m}.
\end{equation}

Finally, for discriminating between pure states with a fidelity of $F$, the minimum type 1 error, $\alpha$, for a given type 2 error, $\beta$, is given by\cite{pereira_analytical_2022}
\begin{equation}
    \alpha = \beta - 2\beta F^2 + F\left(F-2\sqrt{(1-\beta)\beta(1-F^2)}\right).
\end{equation}

\subsection{Quantum ROC}

For a TMSV probe, the initial covariance matrix of the probe (before one mode is sent through the channel) is
\begin{equation}
    V^{\mathrm{in}} = \begin{pmatrix}
    (2m+1)\mathbf{I} &2\sqrt{m(m+1)}\mathbf{Z}\\
    2\sqrt{m(m+1)}\mathbf{Z} &(2m+1)\mathbf{I}
    \end{pmatrix},
\end{equation}
where $m$ is the average photon number of each of the modes. The covariance matrix of the return state is
\begin{equation}
    V^{\mathrm{out}}_{(i)} = \begin{pmatrix}
    (2m+1)\mathbf{I} &2\sqrt{\tau_i m(m+1)}\mathbf{Z}\\
    2\sqrt{\tau_i m(m+1)}\mathbf{Z} &(2m\tau+1)\mathbf{I}
    \end{pmatrix},
\end{equation}
where $i$ depends on the identity of the channel. This covariance matrix can be diagonalized by a two-mode squeezing (unitary) operation, with the squeezing parameter depending on $\tau_i$. The resulting state is the tensor product of a vacuum state and a thermal state, with average photon number $\bar{n}_i$ (which again depends on $\tau_i$). Let us call the channel output $\rho'_i$, and let us call the diagonalizing unitary for that output state $U_i$. Then,
\begin{equation}
    \begin{split}
        &U_i \rho'_i U_i^{\dagger} = \left|0\middle>\middle<0\right|\otimes \sigma_i,\\
        &\sigma_i = \sum_{n=0}^{\infty} \frac{\bar{n}_i^n}{(\bar{n}_i+1)^{n+1}} \left|n\middle>\middle<n\right|.
    \end{split}\label{eq: diagonal state}
\end{equation}
Consequently, we can apply $U_1$ to our output state, defining $\rho_i = U_1 \rho'_i U_1^{\dagger}$, such that
\begin{equation}
    \begin{split}
        &\rho_1 = \left|0\middle>\middle<0\right|\otimes \sigma_1,\\
        &\rho_2 = U^{\dagger}(\left|0\middle>\middle<0\right|\otimes \sigma_2)U,
        ~~U=U_2 U_1^{\dagger}.
    \end{split}
\end{equation}
Note that, as a composition of two-mode squeezing operations, $U$ is also a two-mode squeezing operation, and so can be written as~\cite{caves_new_1985}
\begin{equation}
    U(r) = \exp \left[r(\hat{a}\hat{b}-\hat{a}^{\dagger}\hat{b}^{\dagger})\right].\label{eq: U expression}
\end{equation}

Now, suppose we carried out a photon counting measurement on the first mode of state $\rho_i$. If $i=1$, we always get a result of $0$ (a type 1 error of $0$). On the other hand, if $i=2$, we get a result of $0$ with some non-zero probability (a non-zero type 2 error). This is one possible measurement scheme.

Suppose, on the other hand, we carried out a photon counting measurement on the first mode of state $U \rho_i U^{\dagger}$ (i.e. apply the two-mode squeezing unitary, $U$, prior to the measurement). Now, if $i=1$, we get a result of $0$ some of the time, but if $i=2$, we always get a result of $0$. In this case, therefore, we get a non-zero type 1 error and a type 2 error of 0. This defines another possible measurement scheme. Note that in both schemes, we only care about whether the result is $0$ or not $0$ (not the actual number), so this is more akin to a click detector.

On a plot of type 1 error against type 2 error (ROC), we can draw a straight line between these two points, and achieve any pair of errors along this line. This could be achieved by carrying out the first measurement with probability $a$ and the second measurement with probability $1-a$. Suppose we want a better than linear interpolation between the two points. How might we go about designing a measurement to achieve this?

One thing we might consider is, instead of choosing one measurement or the other with some classical probability, we could control which measurement is carried out using a quantum state. Suppose we apply a controlled unitary to the state $\rho_i$, so that if the control qubit is $\left|0\right>$, we apply the identity to $\rho_i$ and if the control qubit is $\left|1\right>$, we apply $U$. Let us denote the resulting channel as $\mathcal{U}$, and write
\begin{equation}
    \begin{split}
        \mathcal{U}:& \left|a\middle>\middle<a\right|\otimes \rho_i \to a\left|0\middle>\middle<0\right|\otimes \rho\\
        &+ \sqrt{a(1-a)}\left(\left|0\middle>\middle<1\right|\otimes \rho_i U^{\dagger} + \left|1\middle>\middle<0\right|\otimes U\rho_i\right)\\
        &+ (1-a) \left|1\middle>\middle<1\right|\otimes U\rho U^{\dagger},
    \end{split}
\end{equation}
where $\left|a\right> = \sqrt{a}\left|0\right> + \sqrt{1-a}\left|1\right>$ is a control qubit.

If we now measured the control qubit and then the first mode, this would reduce to the classical combination of measurements. But by retaining the off-diagonal components, we may be able to do better. To reduce the complexity of the problem, let us apply the following channel to the return state (and the identity to the control qubit), retaining the superposition of measurements:
\begin{align}
    &\mathcal{C}_{\mathrm{CV}\to\mathrm{DV}}[\rho] = \sum_{k} K_k \rho K_k^{\dagger},\\
    &\{K_k\} = \{\left|00\middle>\middle<00\right|,\left|01\middle>\middle<0i\right|,\left|10\middle>\middle<i0\right|,\left|11\middle>\middle<ij\right|\},
\end{align}
where $i$ and $j$ both run from $1$ to $\infty$. In other words, this channel, when applied to a two-mode Gaussian state, maps it to a two-qubit state by mapping all of the non-vacuum components to the single outcome $\left|1\right>$. Note that, by doing this, we lose some information, but not all of it, and that we have not traced over the second mode, even though the original two protocols trace over it (by only carrying out photon counting on the first mode). The reason we do not simply trace over the second mode is to retain as much information as possible, whilst still having a state that is easier to work with.

Let us calculate the resulting three-qubit states, $\rho_i^{\mathrm{out}}=(\mathbf{I}\otimes\mathcal{C}_{\mathrm{CV}\to\mathrm{DV}})\cdot\mathcal{U}[\left|a\middle>\middle<a\right|\otimes \rho_i]$. We will do this component by component and state by state. Let us start with 
\begin{equation}
    \begin{split}
    \mathcal{C}_{\mathrm{CV}\to\mathrm{DV}}[\rho_1] =& \left<0\middle|\sigma_1\middle|0\right>\left|00\middle>\middle<00\right|\\
    &+(1-\left<0\middle|\sigma_1\middle|0\right>)\left|01\middle>\middle<01\right|\\
    =& x_1\left|00\middle>\middle<00\right|+(1-x_1)\left|01\middle>\middle<01\right|,
    \end{split}
\end{equation}
where we define $x_i = (\bar{n}_i+1)^{-1}$.

Next, we will calculate $\mathcal{C}_{\mathrm{CV}\to\mathrm{DV}}[U\rho_1]$, using the Fock basis expansion of $\sigma_1$ in Eq.~(\ref{eq: diagonal state}).
\begin{equation}
    \mathcal{C}_{\mathrm{CV}\to\mathrm{DV}}[U\rho_1] = \mathcal{C}_{\mathrm{CV}\to\mathrm{DV}}\left[\sum_{m=0}^{\infty} \frac{\bar{n}_1^m U\left|0m\middle>\middle<0m\right|}{(\bar{n}_1+1)^{m+1}} \right].
\end{equation}
The key quantity we need to calculate here is
\begin{equation}
    \left<i,j\middle|U\middle|0,m\middle>\middle<0,m\middle|i,j\right> = \delta_{i0}\delta_{jm}\left<0,m\middle|U\middle|0,m\right>.
\end{equation}
We can decompose $U$ as~\cite{schumaker_new_1985}
\begin{equation}
    \begin{split}
        U(r) =& \cosh(r)^{-1}\exp\left[-\tanh(r)\hat{a}^{\dagger}\hat{b}^{\dagger}\right]\\
        &\times\exp\left[-\ln(\cosh(r))(\hat{a}^{\dagger}\hat{a}+\hat{b}^{\dagger}\hat{b})\right]\\
        &\times\exp\left[\tanh(r)\hat{a}\hat{b}\right],
    \end{split}
\end{equation}
which is much simpler to use, since we apply the lowering and raising operators in a fixed order, rather than in every possible sequence. The resultant state is~\cite{chizhov_photon_1993}
\begin{equation}
    \begin{split}
        U(r)\left|0,m\right>=&\cosh(r)^{-(m+1)}\sum_{k=0}^{\infty} \Bigg[\binom{k+m}{m}^{\frac{1}{2}}\\
        &\times(-\tanh(r))^{k}\left|k\right>\left|k+m\right>\Bigg].
    \end{split}\label{eq: squeezed Fock state}
\end{equation}
Consequently,
\begin{equation}
    \left<i,j\middle|U\middle|0,m\middle>\middle<0,m\middle|i,j\right> = \delta_{i0}\delta_{jm}\cosh(r)^{-(m+1)},
\end{equation}
and so
\begin{equation}
    \begin{split}
        \mathcal{C}_{\mathrm{CV}\to\mathrm{DV}}[U\rho_1] =& x_1 y\left|00\middle>\middle<00\right|\\
        &+ \frac{(1-x_1)x_1 y^2}{1-(1-x_1)y}\left|01\middle>\middle<01\right|,
    \end{split}
\end{equation}
where $y=\cosh(r)^{-1}$.

We now calculate $\mathcal{C}_{\mathrm{CV}\to\mathrm{DV}}[U\rho_1 U^{\dagger}]$, again making use of Eq.~(\ref{eq: squeezed Fock state}). We can write
\begin{equation}
    \begin{split}
        \left<ij\middle|U\middle|0m\middle>\middle<0m\middle|U^{\dagger}\middle|ij\right> =& \delta_{j-i,m} \cosh(r)^{-2(m+1)}\\
        &\times\tanh(r)^{2i}\binom{i+m}{m}.
    \end{split}\label{eq: coefficients of squeezed Fock}
\end{equation}
Then, setting $i=0$ and $j=0$, we get
\begin{equation}
    \left<00\middle|U \rho U^{\dagger}\middle|00\right>=x_1 y^2.
\end{equation}
Setting $i=1$ and instead summing over $j$, we get
\begin{equation}
    \sum_{j=1}^{\infty} \left<0j\middle|U \rho U^{\dagger}\middle|0j\right>=\frac{(1-x_1)x_1 y^4}{1-(1-x_1)y^2}.
\end{equation}
Finally, noting that, if $i=0$ and $j\neq 0$, the right-hand side of Eq.~(\ref{eq: coefficients of squeezed Fock}) goes to $0$, we can write
\begin{equation}
    \begin{split}
        \sum_{i,j=1}^{\infty} \left<ij\middle|U \rho U^{\dagger}\middle|ij\right>=& 1 - \left<00\middle|U \rho U^{\dagger}\middle|00\right>\\
        &- \sum_{j=1}^{\infty} \left<0j\middle|U \rho U^{\dagger}\middle|0j\right>\\
        =& \frac{1-y^2}{1-(1-x_1)y^2}.
    \end{split}
\end{equation}
Therefore, our expression for $\mathcal{C}_{\mathrm{CV}\to\mathrm{DV}}[U\rho_1 U^{\dagger}]$ is
\begin{equation}
    \begin{split}
        \mathcal{C}_{\mathrm{CV}\to\mathrm{DV}}[U\rho_1 U^{\dagger}] =& x_1 y^2\left|00\middle>\middle<00\right|\\
        &+ \frac{(1-x_1)x_1 y^4}{1-(1-x_1)y^2}\left|01\middle>\middle<01\right|\\
        &+ \frac{1-y^2}{1-(1-x_1)y^2}\left|11\middle>\middle<11\right|.
    \end{split}
\end{equation}

Putting all of these elements together, we can write the three-qubit state
\begin{equation}
    \begin{split}
        \rho_{1,a}^{\mathrm{out}}=&
        \mathrm{diag}\bigg[a x_1,a(1-x_1),0,0,(1-a)x_1 y^2,\\
        &\frac{(1-a)(1-x_1)x_1 y^4}{1-(1-x_1)y^2},0,\frac{(1-a)(1-y^2)}{1-(1-x_1)y^2}\bigg]\\
        &+\begin{pmatrix}
        \bm{0}_4 &\bm{\omega}_1\\
        \bm{\omega}_1 &\bm{0}_4
        \end{pmatrix}
    \end{split}
\end{equation}
where $\bm{0}_4$ is the $4$ by $4$ zero matrix and we define
\begin{equation}
    \bm{\omega}_i = \begin{pmatrix}
    \sqrt{a(1-a)}x_i y &0 &0 &0\\
    0 &\frac{\sqrt{a(1-a)}(1-x_i)x_i y^2}{1-(1-x_i)y} &0 &0\\
    0 &0 &0 &0\\
    0 &0 &0 &0\\
    \end{pmatrix}.
\end{equation}
Similarly, we can write
\begin{equation}
    \begin{split}
        \rho_{2,a}^{\mathrm{out}}=&
        \mathrm{diag}\bigg[a x_2 y^2,\frac{a(1-x_2)x_2 y^4}{1-(1-x_2)y^2},0,\frac{a(1-y^2)}{1-(1-x_2)y^2},\\
        &(1-a) x_2,(1-a)(1-x_2),0,0\bigg]\\
        &+\begin{pmatrix}
        \bm{0}_4 &\bm{\omega}_2\\
        \bm{\omega}_2 &\bm{0}_4
        \end{pmatrix}
    \end{split}
\end{equation}

Thus, we have reduced a two-mode continuous variable system to a sparse (nine element), three-qubit state, for which the optimal errors can be more easily calculated. For a given value of the parameter $a$, we can calculate optimal values of the type 1 and 2 errors using
\begin{align}
    &\alpha(a,b) = \mathrm{Tr}[(\mathbf{1}-\Pi_{a,b}) \rho_{1,a}^{\mathrm{out}}],\\
    &\beta(a,b) = \mathrm{Tr}[\Pi_{a,b} \rho_{2,a}^{\mathrm{out}}],\\
    &\Pi_{a,b}=\{(1-b)\rho_{2,a}^{\mathrm{out}}-b\rho_{1,a}^{\mathrm{out}}\}_{-},
\end{align}
where $\{X\}_-$ is the projector onto the negative eigenspace of $X$.

Even for our reduced system, it is still difficult to analytically find the ROC by optimizing over parameter $a$ for each value of $b$. Instead, we numerically sample a large number of $a$ and $b$ values to get a large number of pairs $\{\alpha(a,b),\beta(a,b)\}$, then join up the bottom of this set of values to approximate the ROC curve.

\section{Calculation of the mutual information}\label{appendix: MI}

In both scenarios, our aim is to calculate the mutual information between the ground truth (variable $A$) and the estimate of the ground truth (variable $D$). For both scenarios, variables $A$ and $D$ are the same size: in Section~\ref{subsection: k-medoids}, they are pairs of coordinates ($A$ is composed of the true locations of the two attractors and $D$ is composed of the locations of the two cluster centers as found by the clustering algorithm), and in Section~\ref{subsection: dbscan}, they are integers between $0$ and $10$ ($A$ is the number of particles on the surface and $D$ is the number of clusters found by DBSCAN). Note that DBSCAN may find more than $10$ clusters; in this case we replace the actual number of clusters found with $10$, since we know this is the maximum number of particles.

Let us first discuss how we can determine the entropy of a probability distribution over a finite number of outcomes by sampling the distribution a large number of times. If we (independently) sample the distribution a large number of times, we can approximate the true probability distribution using the number of occurrences of each outcome (i.e. the probability of an outcome is approximately the number of occurrences divided by the total number of samples). We call this approximated probability distribution the sample distribution. We can then calculate the entropy of the sample distribution (the sample entropy) in order to approximate the entropy of the true probability distribution. This method is called the plugin estimator.

In the asymptotic limit of $N\gg P$ samples, where $P$ is the number of possible outcomes, the value we calculate will be normally distributed around an expected value. This expected value is not the same as the true value of the entropy, since our estimator is biased: for any finite $N$, the expected value of the plugin estimator is less than the true entropy. We have the following conditions for the variance and bias of the estimator\cite{antos_convergence_2001,basharin_statistical_1959}:
\begin{align}
    &\mathrm{var}(\hat{H}_N) \leq \frac{\mathrm{log}^2_2 N}{N},\label{eq: var}\\
    &\mathbf{E}[\hat{H}_N] = H - \frac{P-1}{2N}\mathrm{log}_2 e + \mathcal{O}[N^{-2}],\label{eq: bias}
\end{align}
where $H$ is the true entropy and $\hat{H}_N$ is the sample entropy for $N$ samples.

We will first consider the scenario in Section~\ref{subsection: dbscan}. Here, we generate $N$ samples according to the flowchart in Fig.~\ref{fig: flowchart}, recording variables $A$ and $D$ for each. We then calculate the sample entropy, $\hat{H}_N(D)$, and the sample conditional entropy, $\hat{H}_N(D|A)$. The conditional entropy is given by
\begin{equation}
    H(D|A) = \sum_a p(A=a)H(D|A=a),
\end{equation}
and so, recalling that $A$ has a uniform probability over all outcomes, we can write
\begin{equation}
    \hat{H}_N(D|A) = \frac{1}{m+1}\sum_{a=0}^m \hat{H}_{\frac{N}{m+1}}(D|A=a),
\end{equation}
where $m$ is the maximum number of clusters. Note that $H(D|A=a)$ is the entropy conditioned on a particular outcome of $A$, and therefore is calculated using only $\frac{1}{m+1}$ of the total number of samples.

The plugin estimator of the conditional entropy is a sum of independent normal distributions and therefore has the following conditions on its variance and bias:
\begin{align}
    &\mathrm{var}(\hat{H}_N(D|A)) \leq \frac{(m+1)\mathrm{log}^2_2 \frac{N}{m+1}}{N},\\
    &\mathbf{E}[\hat{H}_N(D|A)] = H(D|A) - \frac{m(m+1)}{2N}\mathrm{log}_2 e + \mathcal{O}[N^{-2}].
\end{align}
The plugin estimator of the mutual information is given by
\begin{equation}
    \hat{I}_N(A:D) = \hat{H}_N(D) - \hat{H}_N(A|D),
\end{equation}
and so has the following conditions on its variance and bias:
\begin{align}
    &\mathrm{var}(\hat{I}_N(A:D)) \leq \frac{\mathrm{log}^2_2 N + (m+1)\mathrm{log}^2_2 \frac{N}{m+1}}{N},\\
    &\mathbf{E}[\hat{I}_N(A:D)] = I(A:D) + \frac{m^2}{2N}\mathrm{log}_2 e + \mathcal{O}[N^{-2}].
\end{align}
Recalling that $m=10$, and setting the number of samples, $N$, to $20000$, we find that the variance of our estimator is less than or equal to approximately $0.0126$ bits and the bias is approximately $-0.0004$ bits.

Next, let us consider how to calculate the entropy for the scenario in Section~\ref{subsection: k-medoids}. The difficulty in this scenario is that both $A$ and $D$ are sets of $m$ coordinates in a $d$ by $d$ grid. This means that the dimension of each is the number of unique choices of $m$ positions out of $d^2$, which is the binomial coefficient $\binom{d^2}{m}$. For a $20$ by $20$ grid and $2$ clusters, this means that $A$ and $D$ each have $79800$ possible outcomes. This is a lot more than the $11$ possible outcomes for $A$ and $D$ for the scenario in Section~\ref{subsection: dbscan}, and so we need a lot more samples for the plugin estimator, $\hat{H}_N(D)$, to be in the asymptotic regime ($N\gg P$). In order to estimate the conditional entropy in the same way, we would need roughly $80000$ times more samples than even this, because we would be sampling the entropy of $D$ for each possible outcome of $A$.

Instead, we make an assumption about the conditional entropy in order to simplify the calculations. We assume that the entropy of $D$ conditioned on a single value of $A$, $H(D|A=a)$, is approximately the same for all allowed values of $A$. This assumption is justified by the conditions we imposed, which state that the attractors are well separated from each other and are not close to the edge of the grid, so the distributions due to each of the attractors should not interfere with each other.

As a result, we estimate the entropy of $D$ for a small number ($5$) of randomly chosen fixed values of $A$, and use the mean value of these to approximate the conditional entropy, $H(D|A)$. We then assume that the variance of the values we get is approximately the variance of $H(D|A=a)$, i.e. the variance of the entropy of $D$ conditioned on specific values of $A$. Note that the variance of our estimator of $H(D|A=a)$ follows Eq.~(\ref{eq: var}), and so is negligible in the asymptotic regime (we take $800000$ samples, so that $\frac{P}{N}<\frac{1}{10}$).

We estimate $H(D)$ in the normal way, i.e. by randomly choosing values of $A$ for each sample and calculating corresponding values for $D$. Since we use the same number of samples for our estimation of each $H(D|A=a)$ as we do for our estimation of $H(D)$, the biases cancel out, so we have an unbiased estimator of the mutual information, whose variance is approximately the variance of our estimated values of $H(D|A=a)$.

\end{document}